\documentclass[prd,aps, preprintnumbers, showpacs,twocolumn, nofootinbib,superscriptaddress,notitlepage]{revtex4-1}
\usepackage[normalem]{ulem}
\usepackage{epsfig}
\usepackage{amsfonts}
\usepackage{amsmath}
\usepackage{slashed}
\usepackage{graphicx}
\usepackage{color}
\usepackage{mathtools}
\usepackage{subfigure}
\usepackage{stmaryrd}
\usepackage{amssymb,psfrag}
\usepackage{pifont}
\allowdisplaybreaks[4]

\newcommand{\beq}{\begin{eqnarray}}
\newcommand{\eeq}{\end{eqnarray}}

\allowdisplaybreaks

\begin{document}

\title{Direct Light-Cone Reconstruction of the $B$-Meson Shape Function in HQET}

\author{Ao-Sheng Xiong}
\email{xiongash21@lzu.edu.cn}
\affiliation{Frontiers Science Center for RareIsotopes, and School of Nuclear Science and Technology, Lanzhou University, Lanzhou 730000, China}

\author{Shu-Man Hu}
\email{hushm2025@lzu.edu.cn, corresponding author}
\affiliation{Frontiers Science Center for RareIsotopes, and School of Nuclear Science and Technology, Lanzhou University, Lanzhou 730000, China}

\author{Ji Xu}
\email{xuji@lzu.edu.cn, corresponding author}
\affiliation{Frontiers Science Center for RareIsotopes, and School of Nuclear Science and Technology, Lanzhou University, Lanzhou 730000, China}

\author{Fu-Sheng Yu}
\email{yufsh@lzu.edu.cn, corresponding author}
\affiliation{Frontiers Science Center for RareIsotopes, and School of Nuclear Science and Technology, Lanzhou University, Lanzhou 730000, China}
\affiliation{Lanzhou Center for Theoretical Physics, Key Laboratory of Theoretical Physics of Gansu Province, Key Laboratory of Quantum Theory and Applications of MoE, Gansu Provincial Research Center for Basic Disciplines of Quantum Physics, Lanzhou University, Lanzhou 730000, China}

\author{Shuai Zhao}
\email{zhaos@tju.edu.cn, corresponding author}
\affiliation{Department of Physics and Center for Joint Quantum Studies, School of Science, Tianjin University, Tianjin 300350, China}

\begin{abstract}
The $B$-meson shape function encodes the nonperturbative residual-momentum dynamics of the heavy quark within heavy quark effective theory (HQET). From a Wilsonian perspective, HQET provides a natural framework for lattice calculations when the heavy quark mass is comparable to or exceeds the inverse lattice spacing. Motivated by this observation, we introduce a $B$-meson quasi-shape function directly in HQET and calculate its one-loop matching to the light-cone shape function,  avoiding an intermediate matching step involving a full-QCD light-cone distribution. We determine the one-loop renormalization kernels and matching coefficients in both the $\overline{\mathrm{MS}}$ and regularization-independent momentum-subtraction (RI/MOM) schemes for the quasi-shape function, while keeping the light-cone shape function in the $\overline{\mathrm{MS}}$ scheme. The matching coefficients are obtained at leading power in the large-$v^z$ expansion and are independent of the infrared regulator. Our results provide the perturbative ingredients for a direct lattice HQET determination of the $B$-meson shape function.
\end{abstract}

\maketitle

%%%%%%%%%%%%%%%%%%%%%
\section{Introduction}
\label{sec:introduction}
%%%%%%%%%%%%%%%%%%%%%

Precision studies of inclusive $B$-meson decays test the flavor structure of the Standard Model \cite{Shen:2024aba,Blanke:2022deg,BaBar:2004gyj,LHCb:2013ghj}. Interpreting the decay measurements requires a reliable treatment of nonperturbative QCD effects \cite{Beneke:1999br,Khodjamirian:2011ub,Cui:2022zwm}. Nonperturbative effects are particularly important in the endpoint regions of inclusive $B$-meson decays into light hadrons, where a truncated local heavy-quark expansion is insufficient to describe the decay spectra \cite{Bauer:2003pi,Bosch:2004th} as well as the energy flow~\cite{Zhao:2026wuo}. At leading power in the $\Lambda_{\mathrm{QCD}}/m_b$ expansion, the relevant nonperturbative dynamics are encoded in the $B$-meson shape function (SF) within heavy-quark effective theory (HQET) \cite{Neubert:1993ch,Neubert:1993um}. Defined by a forward $B$-meson matrix element of a lightlike bilocal HQET operator, the SF describes the distribution of the heavy quark's residual momentum inside the meson, which is essential theoretical input for the semileptonic decay $B\to X_u\ell\bar\nu$ and the radiative decay $B\to X_s\gamma$. A precise determination of its momentum dependence is therefore important for improving predictions for the endpoint spectra and reducing hadronic uncertainties in extracting the Cabibbo--Kobayashi--Maskawa (CKM) matrix element $|V_{ub}|$ from these semileptonic decay \cite{Cao:2023rku,ParticleDataGroup:2026mpi}.

Previous studies have established the renormalization-scale evolution of the SF and derived relations between its moments and heavy-quark parameters  \cite{Balzereit:1998yf,Neubert:2004sp,Bosch:2004cb}. To describe its nonperturbative momentum dependence, phenomenological models with parametrized functional forms and systematic expansions in basis functions have been employed. The corresponding model parameters or expansion coefficients can then be constrained by theoretical input or data \cite{Ligeti:2008ac}. However, a first-principles determination of this momentum dependence remains challenging. In particular, the lightlike separation in the defining HQET operator prevents direct evaluation of the corresponding matrix element in conventional Euclidean lattice calculations. This motivates relating the light-cone matrix element to a suitable equal-time spacelike correlator accessible on the lattice.

Large-momentum effective theory (LaMET) provides a systematic framework for relating equal-time spatial correlators in boosted hadrons to light-cone distributions \cite{Ji:2013dva,Ji:2014gla,Ji:2020ect}. Progress in applying LaMET and related lattice methods to meson light-cone distribution amplitudes \cite{Xu:2018mpf,Wang:2019msf,Xu:2022krn,LatticeParton:2024zko,LPC:2026vyv,Zhao:2020bsx,Ishaq:2019dst,Zhao:2019elu,Beneke:2023nmj,Izubuchi:2018srq,Xiong:2025obq,Ling:2025olz} motivates extending this strategy to the $B$-meson SF. Within this framework, the Fourier transform of a renormalized spatial correlator in a boosted $B$ meson defines the quasi-SF, which is related to the light-cone SF by a perturbative matching kernel that accounts for the short-distance differences between the spatial and lightlike operators.

In our previous work, we developed a two-step construction of the $B$-meson shape function within heavy-quark large-momentum effective theory (HQLaMET) \cite{Xiong:2026fvs}. A quasi-SF defined with the relativistic QCD $b$-quark field is first matched onto a QCD light-cone SF, which is subsequently matched onto the HQET SF by integrating out the heavy-quark mass scale \cite{Wang:2025uap}. This construction relies on the hierarchy $P_B^z \gg m_b \gg \Lambda_{\mathrm{QCD}}$. However, its implementation on the lattice faces an additional constraint from the finite lattice spacing $a$. A conventional relativistic lattice calculation requires both the heavy-quark mass and the boosted hadron momentum to be sufficiently below the ultraviolet cutoff $1/a$. In particular, the hierarchy $m_b \ll P_B^z \ll 1/a$ becomes difficult to realize, or even parametrically inaccessible, when $m_b$ approaches or exceeds the lattice cutoff.

From an effective field theory perspective, lattice calculations involving bottom quarks can be naturally organized in terms of HQET, particularly when the heavy-quark mass is comparable to or larger than the inverse lattice spacing. In such formulations, the short distance effects associated with the heavy quark mass and the lattice cutoff are incorporated into effective theory matching coefficients, while the lattice calculation resolves the residual momentum dynamics. This motivates constructing the quasi-shape function directly in HQET, so that its spatial to light-cone matching can be performed entirely within the effective theory. In this approach, both operators describe the same low energy heavy quark dynamics, and the large velocity expansion can be performed within HQET itself. Although finite boost and higher order heavy quark corrections remain, this construction avoids the additional QCD-to-HQET matching step and provides a framework naturally adapted to the lattice resolution. We calculate the corresponding renormalization kernels and matching coefficients at one loop, in both the $\overline{\mathrm{MS}}$ and regularization-independent momentum-subtraction (RI/MOM) schemes. Our results provide the perturbative ingredients for a future lattice HQET determination of the $B$-meson shape function, building on earlier developments in lattice calculations with moving heavy quarks \cite{Hashimoto:1995in,Sloan:1997fc,Foley:2002qv,Horgan:2009ti}.

The matching kernel depends on the renormalization scheme chosen for the quasi-distribution \cite{Liu:2018tox,Xu:2022guw}. In this work, we consider the $\overline{\mathrm{MS}}$ and regularization-independent momentum-subtraction (RI/MOM) schemes for the quasi-SF, while keeping the light-cone HQET SF in $\overline{\mathrm{MS}}$ throughout. The $\overline{\mathrm{MS}}$ scheme is standard in continuum perturbation theory, whereas the RI/MOM scheme is well suited to lattice calculations because its renormalization factors can be determined nonperturbatively from off-shell Green functions \cite{Martinelli:1994ty,Lin:2017ani,LatticeParton:2018gjr}. Following the common practice in lattice RI/MOM renormalization, we perform the perturbative calculation of the quasi-SF in Landau gauge.

The remainder of this paper is organized as follows. Section~\ref{sec:definitions} introduces the operator definitions and renormalization prescriptions for the quasi-SF and the light-cone HQET SF, followed by the factorization relation and tree-level matching. Section~\ref{sec:oneloop} presents the bare one-loop partonic matrix elements and derives the corresponding renormalization kernels and matching coefficients in the $\overline{\mathrm{MS}}$ and RI/MOM schemes. Section~\ref{sec:summary} summarizes the main results and outlines prospects for a lattice determination of the HQET SF.

%%%%%%%%%%%%%%%%%%%
\section{$B$-meson shape functions and factorization}
\label{sec:definitions}
%%%%%%%%%%%%%%%%%%%
\subsection{Definitions and notation}

Following the LaMET construction~\cite{Ji:2013dva}, we define the quasi-SF directly in HQET as
\begin{equation}
   \begin{aligned}
      & \widetilde S_{\mathcal R}(\xi,\mu)=\int_{-\infty}^{+\infty}\frac{\mathrm d \tau\,v^z}{2\pi}\,e^{i\xi v^z\tau}\\
      &\qquad \times \frac{\langle B(v)|\bar h_v(0)\,W(0,\tau n_z)\,h_v(\tau n_z)|B(v)\rangle_{\mathcal R}}{\langle B(v)|\bar h_v(0)\,h_v(0)|B(v)\rangle_{\mathcal R}}\,.
   \end{aligned} \label{eq:def_quasi_HQET_SF}
\end{equation}
Here, $h_v$ is the HQET heavy-quark field and $v^\mu=(v^0,0,0,v^z)$ is the $B$-meson four-velocity, with $v^2=1$ and $v^z>0$. We choose the covariant direction components $n_{z\mu}=(0,0,0,1)$, so that $n_z^2=-1$ and $v\cdot n_z=v^z$. The spatial separation is $\tau n_z$, and $\xi$ denotes the residual-momentum conjugate to $v^z\tau$. The straight Wilson line along a direction $n$ is
\begin{equation}
 W(0,\lambda n)=\mathcal P\exp\left[-ig_s\int_0^\lambda\mathrm d s\,n\cdot A(sn)\right]\,,
 \label{eq:wilson}
\end{equation}
where $A$ is the gluon field, $g_s$ is the QCD gauge coupling, and $\mathcal P$ denotes path ordering.

The corresponding light-cone HQET SF is defined by~\cite{Neubert:1993um}
\begin{equation}
   \begin{aligned}
      & S_{\mathcal R}(\omega,\mu)=\int_{-\infty}^{+\infty}\frac{\mathrm d t\,v^+}{2\pi}\,e^{i\omega t v^+}\\
      & \qquad \times\frac{\langle B(v)|\bar h_v(0)\,W(0,tn_+)\,h_v(tn_+)|B(v)\rangle_{\mathcal R}}{\langle B(v)|\bar h_v(0)\,h_v(0)|B(v)\rangle_{\mathcal R}}\,.
   \end{aligned}\label{eq:def_HQET_SF}
\end{equation}
We use $n_{\pm\mu}=(1,0,0,\pm1)/\sqrt2$ and define $v^\pm=n_\pm\cdot v$ and $k^\pm=n_\pm\cdot k$. Thus $n_\pm^2=0$, and the separation $t n_+$ is lightlike. The variable $\omega$ is Fourier conjugate to $v^+t$. Both $\xi$ and $\omega$ have momentum dimension. The two SFs share the same HQET fields and local normalization but differ in their separation directions.

The subscript $\mathcal R$ labels the renormalization
scheme. Throughout this work, the light-cone SF is kept in $\overline{\mathrm{MS}}$ and denoted by $S_{\overline{\mathrm{MS}}}(\omega,\mu)$. On the quasi side we use $\widetilde S_{\overline{\mathrm{MS}}}(\xi,\mu)$ and $\widetilde S_{\mathrm{RI}}(\xi,\{\widetilde\mu\})$, where $\mu$ is the renormalization scale and $\{\widetilde\mu\}$ denotes the RI/MOM reference scales which will be specified below.

\subsection{Bare partonic distribution and local normalization}
\label{sec:baredefinitions}

For the perturbative calculation, the hadronic states in Eqs.~\eqref{eq:def_quasi_HQET_SF} and~\eqref{eq:def_HQET_SF} are replaced by HQET heavy-quark external states with equal incoming and outgoing residual momentum $k$. We denote the bare spatial bilocal and local matrix elements by $\widetilde{h}_{\mathcal B}(\tau;k)$ and $\widetilde{h}_{\mathcal B}(0;k)$, respectively, with the subscript $\mathcal B$ indicating bare quantities. The locally normalized bare quasi-SF is
\begin{equation}
    \widetilde S_{\mathcal B}(\xi,\mu;k)=\int_{-\infty}^{+\infty}\frac{\mathrm d\tau\,v^z}{2\pi}\,e^{i\xi v^z\tau}\frac{\widetilde{h}_{\mathcal B}(\tau;k)}{\widetilde{h}_{\mathcal B}(0;k)}\,.
\label{eq:barequasi}
\end{equation}
The bare light-cone SF $S_{\mathcal B}(\omega,\mu;k)$ is defined analogously. 

We introduce the residual-momentum label $\widetilde k\equiv k^z/v^z$ and use $k^+/v^+\simeq\widetilde k$ at leading power in the large-boost expansion. At tree level, the spatial matrix-element ratio is $e^{-ik^z\tau}$, yielding the quasi-SF tree-level distribution. The analogous light-cone calculation gives
\begin{equation}
   \widetilde S^{(0)}(\xi)=\delta(\xi-\widetilde k),\qquad S^{(0)}(\omega)=\delta(\omega-\widetilde k)\,.
 \label{eq:tree}
\end{equation}

Including the one-loop corrections, we write the bare and renormalized partonic SFs as
\begin{equation}
\begin{aligned}
 \widetilde S_X(\xi)&=\widetilde S^{(0)}(\xi)+\widetilde S_X^{(1)}(\xi)+\mathcal{O}(\alpha_s^2)\,,\\
 S_X(\omega)&=S^{(0)}(\omega)+S_X^{(1)}(\omega)+\mathcal{O}(\alpha_s^2)\,.
\end{aligned}
\label{eq:SFexpansion}
\end{equation}
Here, $X=\mathcal B$ labels the bare distributions, whereas $X=\mathcal R$ labels the distributions renormalized in scheme $\mathcal R$. The superscripts $(0)$ and $(1)$ denote the tree-level terms and one-loop corrections, respectively, with the factors of $\alpha_s$ included in the latter. For brevity, only the distribution variables $\xi$ and $\omega$ are displayed explicitly in Eq.~\eqref{eq:SFexpansion}.

To obtain the one-loop contributions to the normalized bare distributions, we also expand the bare matrix elements as
\begin{equation}
    \begin{aligned}
         \widetilde h_{\mathcal B}(\tau;k)=\widetilde h^{(0)}(\tau;k)+\widetilde h_{\mathcal B}^{(1)}(\tau;k)+\mathcal{O}(\alpha_s^2)\,,
    \end{aligned}
\end{equation}
with the corresponding expansion for the local matrix element $\widetilde h_{\mathcal B}(0;k)$. Using the tree-level relation, we obtain
\begin{equation}
   \begin{aligned}
      \left(\frac{\widetilde{h}_{\mathcal B}(\tau;k)}{\widetilde{h}_{\mathcal B}(0;k)}\right)^{(1)} =\frac{\widetilde{h}_{\mathcal B}^{(1)}(\tau;k)}{\widetilde{h}^{(0)}(0;k)}-e^{-ik^z\tau}\frac{\widetilde{h}_{\mathcal B}^{(1)}(0;k)}{\widetilde{h}^{(0)}(0;k)}\,.
   \end{aligned} \label{eq:ratioexpand}
\end{equation}
Thus the local correction enters the normalized distribution with a minus sign. We label the four one-loop diagram topologies in Fig.~\ref{fig:feynman} by $(a)$--$(d)$, corresponding to the heavy-quark sail, Wilson-line self-energy, nonlocal box, and local vertex, respectively. The notation $(1,j)$ denotes the one-loop contribution from topology $(j)$, before applying the minus sign associated with the local denominator correction. The normalized bare distributions are therefore assembled as
\begin{equation}
\begin{aligned}
 S_{\mathcal B}^{(1)}&=S_{\mathcal B}^{(1,a)}+S_{\mathcal B}^{(1,b)}+S_{\mathcal B}^{(1,c)}-S_{\mathcal B}^{(1,d)}\,,\\
 \widetilde S_{\mathcal B}^{(1)}&=\widetilde S_{\mathcal B}^{(1,a)}+\widetilde S_{\mathcal B}^{(1,b)}+\widetilde S_{\mathcal B}^{(1,c)}-\widetilde S_{\mathcal B}^{(1,d)}\,.
\end{aligned}
\label{eq:bareassembly}
\end{equation}
The common external-leg factors cancel between the bilocal and local matrix elements and therefore do not contribute to the normalized ratio.

For the one-loop calculation, we use dimensional regularization $d=4-2\epsilon$, with $\mu$ denoting the $\overline{\mathrm{MS}}$ renormalization scale. The nonzero residual energy $v\cdot k$ is retained as a common infrared regulator in both partonic SFs. The endpoint singularities are treated using the generalized plus prescription
\begin{equation}
   [F(\omega,\widetilde k)]_{\oplus}=F(\omega,\widetilde k)-\delta(\omega-\widetilde k)\int_0^{a\omega}\mathrm d t\,F(\omega,t)\,,
\label{eq:plus}
\end{equation}
where $a>1$ is a common auxiliary subtraction parameter. For the quasi-SF, the same prescription applies with $\omega$ replaced by $\xi$.

\subsection{Renormalization prescriptions}
\label{sec:prescriptions}

Since both SFs are defined from ratios of bilocal and local matrix elements, their renormalization involves the factors associated with both operators. In coordinate space, we adopt the convention
\begin{equation}
 \langle O\rangle_{\mathcal B} = Z_{\mathcal R}\langle O\rangle_{\mathcal R}\,,
\label{eq:opren}
\end{equation}
with a separate factor for each operator. We denote the factors for the spatial and lightlike bilocal operators by $Z_{\mathcal R}(\tau)$ and $Z_{\mathcal R}^{\mathrm{LC}}(t)$, respectively. The common local normalization operator is assigned the factor $Z_{\mathcal R}(0)$ in both definitions.

We denote the spatial and light-cone bilocal matrix elements by $\widetilde h_X(\tau;k)$ and $h_X(t;k)$, respectively, and their local counterparts by $\widetilde h_X(0;k)$ and $h_X(0;k)$, where $X=\mathcal B$ or $\mathcal R$ labels bare or renormalized quantities, respectively.

The renormalization relation for the normalized spatial matrix elements follows from the multiplicative renormalization of the bilocal and local operators:
\begin{equation}
 \frac{\widetilde h_{\mathcal B}(\tau;k)}{\widetilde h_{\mathcal B}(0;k)}=\frac{Z_{\mathcal R}(\tau)}{Z_{\mathcal R}(0)}\frac{\widetilde h_{\mathcal R}(\tau;k)}{\widetilde h_{\mathcal R}(0;k)}\,.
\label{eq:ratioren}
\end{equation}
The light-cone matrix elements obey an analogous relation,
\begin{equation}
   \frac{h_{\mathcal B}(t;k)}{h_{\mathcal B}(0;k)}=\frac{Z_{\mathcal R}^{\mathrm{LC}}(t)}{Z_{\mathcal R}(0)}\frac{h_{\mathcal R}(t;k)}{h_{\mathcal R}(0;k)}\,.
 \label{eq:ratiorenLC}
\end{equation}

Fourier transforming the coordinate-space relations in Eqs.~\eqref{eq:ratioren} and~\eqref{eq:ratiorenLC} gives the following convolution relations in momentum space:
\begin{equation}
   \begin{aligned}
      \widetilde S_{\mathcal B}(\xi)&=\int \mathrm d\xi'\,\widetilde{\mathcal Z}_{\mathcal R}(\xi,\xi')\widetilde S_{\mathcal R}(\xi'),\\
      S_{\mathcal B}(\omega)&=\int \mathrm d\omega'\,\mathcal Z_{\mathcal R}(\omega,\omega') S_{\mathcal R}(\omega')\,.
   \end{aligned}\label{eq:SFrenconvolution}
\end{equation}
Here, $\widetilde{\mathcal Z}_{\mathcal R}$ and $\mathcal Z_{\mathcal R}$ denote the momentum-space renormalization kernels for the quasi-SF and light-cone SF, respectively. The primed variables are integrated over, while $\xi$ and $\omega$ label the bare distributions on the left-hand side. The kernels are the Fourier transforms of the corresponding ratios of renormalization factors in coordinate space:
\begin{equation}
   \widetilde{\mathcal Z}_{\mathcal R}(\xi,\xi')=\int_{-\infty}^{+\infty}\frac{\mathrm d\tau\,v^z}{2\pi}\,e^{i(\xi-\xi')v^z\tau}\frac{Z_{\mathcal R}(\tau)}{Z_{\mathcal R}(0)}\,,
\label{eq:kernelFourier}
\end{equation}
and
\begin{equation}
    \mathcal Z_{\mathcal R}(\omega,\omega')=\int_{-\infty}^{+\infty}\frac{\mathrm dt\,v^+}{2\pi}\,e^{i(\omega-\omega')v^+t}\frac{Z_{\mathcal R}^{\mathrm{LC}}(t)}{Z_{\mathcal R}(0)}\,.
\label{eq:kernelFourierLC}
\end{equation}
In the following, the light-cone SF is renormalized in $\overline{\mathrm{MS}}$, whereas the quasi-SF is considered in both the $\overline{\mathrm{MS}}$ and RI/MOM schemes.

At tree level, the ratios of renormalization factors are unity, so both kernels reduce to delta distributions. Through one loop, we write
\begin{equation}
   \begin{aligned}
      \widetilde{\mathcal Z}_{\mathcal R}(\xi,\xi')&=\delta(\xi-\xi')+\widetilde{\mathcal Z}_{\mathcal R}^{(1)}(\xi,\xi')+\mathcal{O}(\alpha_s^2)\,,\\
      \mathcal Z_{\mathcal R}(\omega,\omega')&=\delta(\omega-\omega')+\mathcal Z_{\mathcal R}^{(1)}(\omega,\omega')+\mathcal{O}(\alpha_s^2)\,.
\end{aligned}
\label{eq:Zexpansion}
\end{equation}

The tree-level delta distributions in Eq.~\eqref{eq:tree} therefore fix the primed integration variables to $\widetilde k$. Expanding Eq.~\eqref{eq:SFrenconvolution} then gives
\begin{equation}
\begin{aligned}
   \widetilde S_{\mathcal R}^{(1)}(\xi) &= \left. \widetilde S_{\mathcal B}^{(1)}(\xi)-\widetilde{\mathcal Z}_{\mathcal R}^{(1)}(\xi,\xi') \right|_{\xi' \to \widetilde k} \,,\\
   S_{\mathcal R}^{(1)}(\omega)&= \left. S_{\mathcal B}^{(1)}(\omega)-\mathcal Z_{\mathcal R}^{(1)}(\omega,\omega') \right|_{\omega' \to \widetilde k}\,.
\end{aligned}
\label{eq:commonsubtraction}
\end{equation}
The prescriptions below determine these kernels from the bare partonic matrix elements. The renormalized one-loop SFs then follow from the subtraction relations in Eq.~\eqref{eq:commonsubtraction}.

\subsubsection{$\overline{MS}$ prescription}

The one-loop $\overline{\mathrm{MS}}$ kernels are determined by the ultraviolet pole parts of the normalized bare distributions, with no additional finite subtraction. Denoting this pole extraction in the $\overline{\mathrm{MS}}$ convention by $[\ldots]_{\mathrm{UV}}$, we obtain
\begin{equation}
\begin{aligned}
   \widetilde{\mathcal Z}_{\overline{\mathrm{MS}}}^{(1)}(\xi,\xi')&=\left.\big[\widetilde S_{\mathcal B}^{(1)}(\xi,\widetilde{k})\big]_{\mathrm{UV}}\right|_{\widetilde k\to\xi'}\,,\\
   \mathcal Z_{\overline{\mathrm{MS}}}^{(1)}(\omega,\omega')&=\left.\big[S_{\mathcal B}^{(1)}(\omega,\widetilde{k})\big]_{\mathrm{UV}}\right|_{\widetilde k\to\omega'}\,.\\
\end{aligned}
\label{eq:MSproject}
\end{equation}

\subsubsection{RI/MOM prescription}
The RI/MOM condition also specifies the finite subtraction on the quasi side. We choose an off-shell reference momentum $k_R^\mu=(k_R^0,0,0,k_R^z)$ and define
\begin{equation}
\begin{aligned}
 \{\widetilde\mu\}&\equiv\{k_R^2,~ k_R^z\}\,,&
 r_R&\equiv\frac{k_R^2}{(k_R^z)^2}\,,\\
 k_R^0&=k_R^z\sqrt{1+r_R}\,,&
 \widetilde k_R&\equiv\frac{k_R^z}{v^z}\,.
\end{aligned}
\label{eq:reference}
\end{equation}
We use the same fixed values of $k_R^2$ and $k_R^z$ to renormalize the quasi-SF at different external momenta $k$.

We impose the RI/MOM condition on off-shell heavy-quark Green functions evaluated in Landau gauge. At $k=k_R$, the renormalized ratio is required to equal its tree-level value:
\begin{equation}
 \frac{Z_{\mathrm{RI}}^{-1}(\tau,\{\widetilde\mu\})\widetilde{h}_{\mathcal B}(\tau;k_R)}
 {Z_{\mathrm{RI}}^{-1}(0,\{\widetilde\mu\})\widetilde{h}_{\mathcal B}(0;k_R)}
 =e^{-ik_R^z\tau}\,.
 \label{eq:RIcondition}
\end{equation}
Using Eq.~\eqref{eq:ratioren}, this condition determines the ratio of renormalization factors directly from the bare reference-point matrix elements,
\begin{equation}
\frac{Z_{\mathrm{RI}}(\tau,\{\widetilde\mu\})}{Z_{\mathrm{RI}}(0,\{\widetilde\mu\})}
 =e^{ik_R^z\tau}\frac{\widetilde{h}_{\mathcal B}(\tau;k_R)}{\widetilde{h}_{\mathcal B}(0;k_R)}\,.
 \label{eq:RIfactor}
\end{equation}

We denote the one-loop contribution to the normalized bare quasi-SF at $k=k_R$ by $\widetilde S_{\mathcal B,\mathrm{CT}}^{(1)}$. Its inverse Fourier transform is
\begin{equation}
 \left(\frac{\widetilde{h}_{\mathcal B}(\tau;k_R)}{\widetilde{h}_{\mathcal B}(0;k_R)}\right)^{(1)}
 =\int\mathrm d\eta\,e^{-iv^z\eta\tau}\widetilde S_{\mathcal B,\mathrm{CT}}^{(1)}(\eta,\{\widetilde\mu\})\,.
 \label{eq:CTdef}
\end{equation}
The label CT identifies the counterterm contribution that defines the subtraction. It includes both the ultraviolet poles and the finite reference-point terms. Its diagram contributions are combined according to Eq.~\eqref{eq:bareassembly}, including the minus sign from the local denominator correction. The explicit reference-point results in Sec.~\ref{sec:RI_renormalization} are expanded at large $v^z$ with $k_R^z$ and $r_R$ held fixed.

Expanding Eq.~\eqref{eq:RIfactor} through one loop and using Eq.~\eqref{eq:CTdef}, we find
\begin{equation}
   \begin{aligned}
      & \left(\frac{Z_{\mathrm{RI}}(\tau,\{\widetilde\mu\})}{Z_{\mathrm{RI}}(0,\{\widetilde\mu\})}\right)^{(1)} = \int\mathrm d\eta\,e^{-iv^z(\eta-\widetilde k_R)\tau}\widetilde S_{\mathcal B,\mathrm{CT}}^{(1)}(\eta,\{\widetilde\mu\})\,.
   \end{aligned}\label{eq:RIfactorone}
\end{equation}
The reference phase $e^{ik_R^z\tau}$ shifts the Fourier argument. Substituting Eq.~\eqref{eq:RIfactorone} into Eq.~\eqref{eq:kernelFourier} therefore gives the one-loop renormalization kernel
\begin{equation}
   \begin{aligned}
      \widetilde{\mathcal Z}_{\mathrm{RI}}^{(1)}(\xi,\xi';\{\widetilde\mu\}) & = \left.\widetilde S_{\mathcal B,\mathrm{CT}}^{(1)}\bigl(\xi+\widetilde k_R-\widetilde k,\{\widetilde\mu\}\bigr)\right|_{\widetilde k\to\xi'}\\
      &=\widetilde S_{\mathcal B,\mathrm{CT}}^{(1)}\bigl(\xi+\widetilde k_R-\xi',\{\widetilde\mu\}\bigr)\,.
   \end{aligned}\label{eq:RIshiftkernel}
\end{equation}

\subsection{Factorization and perturbative matching}
\label{sec:factorization}

At leading power in the large-boost expansion, the renormalized quasi-SF is related to the $\overline{\mathrm{MS}}$ light-cone HQET SF through

\begin{equation}
 \widetilde S_{\mathcal R}(\xi)=\int\mathrm d\omega\,
 H_{\mathcal R/\overline{\mathrm{MS}}}(\xi,\omega,v^z;\mu)\,S_{\overline{\mathrm{MS}}}(\omega,\mu)\,.
 \label{eq:factorization}
\end{equation}
The integral extends over the support of the light-cone SF; power corrections are beyond the accuracy considered here. The first and second scheme labels on $H_{\mathcal R/\overline{\mathrm{MS}}}$ specify the quasi- and light-cone prescriptions. We consider $H_{\overline{\mathrm{MS}}/\overline{\mathrm{MS}}}$ and $H_{\mathrm{RI}/\overline{\mathrm{MS}}}$, where RI abbreviates RI/MOM only in the matching-kernel label. A common scale $\mu$ is used for the $\overline{\mathrm{MS}}$ quantities, while $H_{\mathrm{RI}/\overline{\mathrm{MS}}}$ also depends on $\{\widetilde\mu\}$.

The matching coefficients are determined from the partonic SFs renormalized in their respective schemes and evaluated with the same external state and infrared regulator. Together with the SF expansions in Eq.~\eqref{eq:SFexpansion}, we use
\begin{equation}
 H_{\mathcal R/\overline{\mathrm{MS}}}(\xi,\omega)=H^{(0)}(\xi,\omega)+H_{\mathcal R/\overline{\mathrm{MS}}}^{(1)}(\xi,\omega)+\mathcal{O}(\alpha_s^2)\,.
 \label{eq:Hexpansion}
\end{equation}
The tree-level distributions in Eq.~\eqref{eq:tree} give
\begin{equation}
 H^{(0)}(\xi,\omega)=\delta(\xi-\omega)\,.
 \label{eq:Htree}
\end{equation}

At one loop, expanding the convolution in Eq.~\eqref{eq:factorization} gives
\begin{equation}
 H_{\mathcal R/\overline{\mathrm{MS}}}^{(1)}(\xi,\omega)= \left[\widetilde S_{\mathcal R}^{(1)}(\xi)-\left.S_{\overline{\mathrm{MS}}}^{(1)}(\omega)\right|_{\omega\to\xi}\right]_{\widetilde k\to\omega}\,.
\label{eq:matching_one_loop}
\end{equation}

\begin{widetext}
%%%%%%%%%%%%%%%%%%%%%	
\section{One-loop renormalization and matching}
\label{sec:oneloop}
%%%%%%%%%%%%%%%%%%%%%

%%%%%%%%%%%%%%%%%%%%%
\subsection{Bare partonic matrix elements}
\label{sec:bare_matrix_elements}
%%%%%%%%%%%%%%%%%%%%%

We evaluate the one-loop diagrams in Fig.~\ref{fig:feynman} for the light-cone and quasi-SFs, using the infrared regulator and generalized plus prescription specified in Sec.~\ref{sec:baredefinitions}. We first present the individual bare diagram contributions, which are then combined according to Eq.~\eqref{eq:bareassembly} to obtain the normalized bare distributions.

Since the light-cone SF is gauge invariant, we calculate it in Feynman gauge for convenience. As a cross-check, we have repeated the calculation in Landau gauge and verified that the complete one-loop results agree.

\begin{figure}[htbp]
  \centering
   \includegraphics[width=0.9\textwidth]{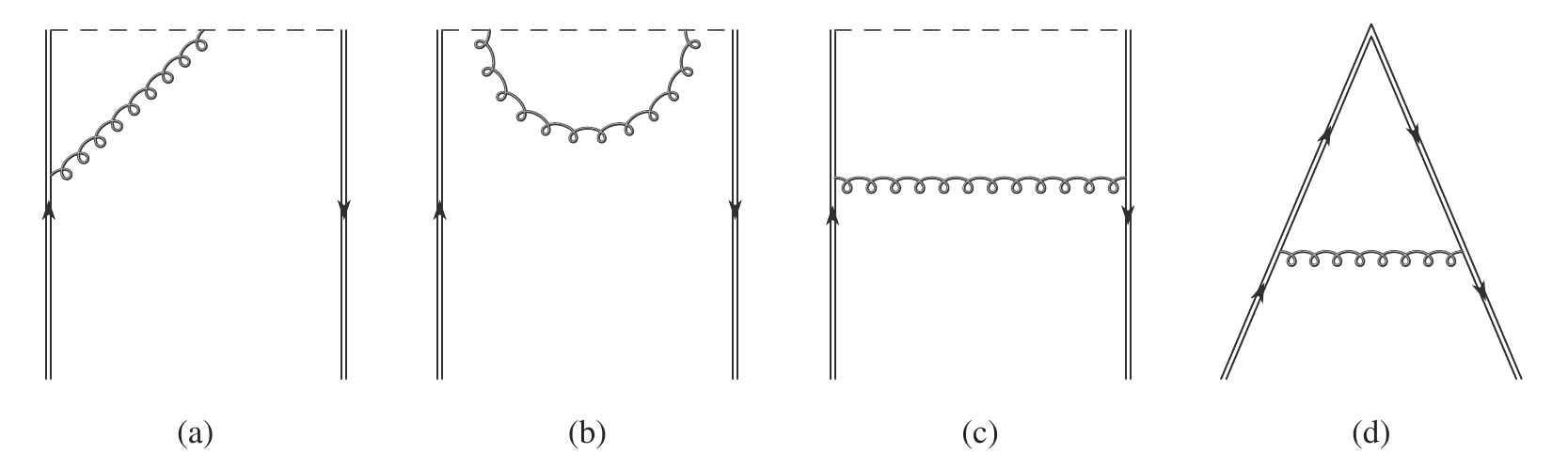}
  \caption{One-loop diagrams for the $B$-meson SFs: (a) heavy-quark sail, (b) Wilson-line self-energy, (c) nonlocal box, and (d) local vertex. Solid and dotted lines denote HQET heavy quarks and Wilson lines, respectively.}
  \label{fig:feynman}
\end{figure}

We first consider the light-cone SF. The sail diagram in Fig.~\ref{fig:feynman}(a) gives

\begin{equation}
    \begin{aligned}
        S^{(1,a)}_{\mathcal{B}}(\omega,\mu) & =\left[\frac{\alpha_s C_F}{2\pi}\left( \frac{2}{\epsilon} \frac{1}{\widetilde{k}-\omega} +  \frac{2}{\widetilde{k}-\omega}\ln \frac{\mu^2}{(\widetilde{k}-\omega)^2} \right) \theta(\widetilde{k}-\omega)\right]_\oplus \\
        &  + \frac{\alpha_s C_F}{2\pi} \delta(\omega-\widetilde{k}) \left( -\frac{1}{\epsilon^2} - \frac{1}{\epsilon} \ln\frac{\mu^2}{(a-1)^2\omega^2} - \frac{1}{2}\ln^2 \frac{\mu^2}{(a-1)^2\omega^2}-\frac{\pi^2}{12} \right)\,.
    \end{aligned}
\end{equation}

The Wilson-line self-energy diagram in Fig.~\ref{fig:feynman}(b) gives
\begin{equation}
   S_{\mathcal{B}}^{(1,b)}(\omega,\mu)=0\,.
\end{equation}
since its numerator is proportional to $n_+^2$, which vanishes for the lightlike link.

The nonlocal box diagram in Fig.~\ref{fig:feynman}(c) yields
\begin{equation}
    \begin{aligned}
        S^{(1,c)}_{\mathcal{B}}(\omega,\mu) =\frac{\alpha_s C_F}{2\pi} \Bigg\{ \left[- \frac{2\theta(\widetilde{k}-\omega)}{\widetilde{k}-\omega}\right]_\oplus- \delta(\omega-\widetilde{k})\ln\frac{(a-1)^2\omega^2}{(-2v\cdot k)^2} \Bigg\}\,.
    \end{aligned}
\end{equation}

The local vertex diagram in Fig.~\ref{fig:feynman}(d) gives
\begin{equation}
    \begin{aligned}
        S^{(1,d)}_{\mathcal{B}}(\omega,\mu) = -\frac{\alpha_s C_F}{2\pi} \delta(\omega-\widetilde{k})\left( \frac{1}{\epsilon} + \ln \frac{\mu^2}{(-2 v\cdot k)^2} \right)\,. 
    \end{aligned}
\end{equation}

We now turn to the bare quasi-SF in Landau gauge. At leading power in the large-$v^z$ expansion, the heavy-quark sail diagram in Fig.~\ref{fig:feynman}(a) gives

\begin{equation}
    \begin{aligned}[b]
        \widetilde{S}^{(1,a)}_{\mathcal{B}}(\xi,\mu) & =\frac{\alpha_s C_F}{2\pi} \Bigg\{ \Bigg[ \frac{\theta(\xi - \widetilde{k})}{\xi - \widetilde{k}} \left( -\frac{1}{\epsilon} + 2\ln \frac{2v^z(\xi - \widetilde{k})}{\mu} + 1 \right) + \theta(\widetilde{k} - \xi) \Bigg( \frac{1}{\widetilde{k} - \xi} \left( \frac{1}{\epsilon} + 2\ln \frac{2\mu v^z}{\widetilde{k} - \xi - 2v \cdot k} \right) \\
        & + \frac{1}{\widetilde{k} - \xi - 2v \cdot k} \Bigg) \Bigg]_{\oplus}  + \delta(\xi - \widetilde{k}) \Bigg[ \left( \ln \frac{a - 1}{4(v^z)^2} - 1 \right) \left( \frac{1}{\epsilon} + \ln \frac{\mu^2}{(a - 1)\xi^2} \right) \\
        & + \frac{1}{2}\ln^2[4(v^z)^2] + \ln[4(v^z)^2] - 2\mathrm{Li}_2\left( \frac{2v \cdot k}{(a - 1)\xi} \right) + \ln\left( 1 - \frac{2v \cdot k}{(a - 1)\xi} \right) + \frac{\pi^2}{6} \Bigg] \Bigg\}\,.
    \end{aligned}\label{eq:quasi_bare_sail}
\end{equation}

The Wilson-line self-energy diagram in Fig.~\ref{fig:feynman}(b) gives
\begin{equation}
    \begin{aligned}
        \widetilde{S}^{(1,b)}_{\mathcal{B}}(\xi,\mu) = \frac{\alpha_s C_F}{4\pi} \Bigg\{ -3 \left[ \frac{\theta(\xi - \widetilde{k})}{\xi - \widetilde{k}} + \frac{\theta(\widetilde{k} - \xi)}{\widetilde{k} - \xi} \right]_{\oplus} + \delta(\xi - \widetilde{k}) \left[ \frac{3}{\epsilon} + 4 + 3 \ln \frac{\mu^2}{4(a - 1)(v^z)^2\xi^2} \right] \Bigg\}\,.
    \end{aligned}\label{eq:quasi_bare_Wilson}
\end{equation}

The nonlocal box diagram in Fig.~\ref{fig:feynman}(c) yields
\begin{equation}
    \begin{aligned}
        \widetilde{S}^{(1,c)}_{\mathcal{B}}(\xi,\mu) & = -\frac{\alpha_s C_F}{4\pi} \Bigg\{ \Bigg[ \frac{\theta(\xi - \widetilde{k})}{\xi - \widetilde{k}} + \frac{4\theta(\widetilde{k} - \xi)}{\widetilde{k} - \xi - 2v \cdot k} + \frac{(\widetilde{k} - \xi)\theta(\widetilde{k} - \xi)}{(\widetilde{k} - \xi - 2v \cdot k)^2} \Bigg]_{\oplus} \\
        & + \delta(\xi - \widetilde{k}) \Bigg[ \ln \frac{2(v^z)^2\xi}{-v \cdot k} + 5\ln\left(1 + \frac{(a - 1)\xi}{-2v \cdot k}\right) - \frac{2v \cdot k}{(a - 1)\xi - 2v \cdot k} - 2 \Bigg] \Bigg\}\,.
    \end{aligned}
\end{equation}

The local vertex diagram in Fig.~\ref{fig:feynman}(d) gives
\begin{equation} 
    \begin{aligned}
        \widetilde{S}^{(1,d)}_{\mathcal{B}}(\xi,\mu) = - \frac{\alpha_s C_F}{4\pi} \delta(\xi - \widetilde{k}) \left[ \frac{3}{\epsilon} + 3 \ln \frac{\mu^2}{4(v \cdot k)^2} - 2 \right]\,.
    \end{aligned} \label{BarequasiSFd}
\end{equation} 

We retain the $v\cdot k$ dependence in the bare quasi-SF results in Eqs.~(\ref{eq:quasi_bare_sail})-(\ref{BarequasiSFd}). At the RI/MOM subtraction point, $v\cdot k$ is replaced by $v\cdot k_R=v^0k_R^0-v^zk_R^z$. With the reference scales held fixed, this contribution cannot be neglected. It determines how the finite terms in the matching coefficient in Eq.~(\ref{eq:matching_kernel_RI}) depend on the RI/MOM subtraction conditions.

%%%%%%%%%%%%%%%%%%%%
\subsection{$\overline{\rm MS}$-to-$\overline{\rm MS}$ matching}
\label{sec:MS_renormalization}
%%%%%%%%%%%%%%%%%%%%
We determine the $\overline{\mathrm{MS}}$ renormalization kernels by applying the ultraviolet-pole extraction prescription in Eq.~\eqref{eq:MSproject}.  For the light-cone SF, the sail diagram supplies the double pole and the nonlocal single pole. This gives
\begin{equation}
   \begin{aligned}
      \mathcal Z_{\overline{\mathrm{MS}}}\left(\omega,\omega';\mu\right) = \delta(\omega-\omega')+\frac{\alpha_s C_F}{2\pi}\Bigg\{\frac{2}{\epsilon}\left[\frac{\theta(\omega'-\omega)}{\omega'-\omega}\right]_\oplus +\delta(\omega-\omega')\left[-\frac{1}{\epsilon^2}+\frac{1}{\epsilon}\left(1-\ln\frac{\mu^2}{(a-1)^2\omega^2}\right)\right]\Bigg\}\,.
    \end{aligned} \label{eq:LC_MS_Z}
\end{equation}
For the quasi-SF, extracting the ultraviolet pole part of the normalized bare distribution according to Eq.~\eqref{eq:MSproject} yields the $\overline{\mathrm{MS}}$ renormalization kernel,
\begin{equation}
   \begin{aligned}
      \widetilde{\mathcal Z}_{\overline{\mathrm{MS}}}\left(\xi,\xi';v^z,\mu\right)=\delta(\xi-\xi') + \frac{\alpha_s C_F}{2\pi\epsilon}\Bigg\{\Bigg[\frac{\theta(\xi'-\xi)}{\xi'-\xi} -\frac{\theta(\xi-\xi')}{\xi-\xi'}\Bigg]_\oplus +\delta(\xi-\xi')\left[\ln\frac{a-1}{4(v^z)^2}+2\right]\Bigg\}\,.
   \end{aligned}\label{eq:quasi_MS_Z}
\end{equation}

With these kernels fixed, Eq.~\eqref{eq:commonsubtraction} gives the renormalized partonic SFs. For the light-cone SF, the result through one loop is
\begin{equation}
   \begin{aligned}
      S_{\overline{\mathrm{MS}}}(\omega,\mu)& =\delta(\omega-\widetilde k)+\frac{\alpha_s C_F}{2\pi}\Bigg\{\Bigg[\frac{2\theta(\widetilde k-\omega)}{\widetilde k-\omega}\ln\frac{\mu^2}{(\widetilde k-\omega)^2}-\frac{2\theta(\widetilde k-\omega)}{\widetilde k-\omega}\Bigg]_{\oplus}\\
      &+\delta(\omega-\widetilde k)\Bigg[-\frac12\ln^2\frac{\mu^2}{(a-1)^2\omega^2}+\ln\frac{\mu^2}{(a-1)^2\omega^2}-\frac{\pi^2}{12}\Bigg]\Bigg\}\,.
\end{aligned}
\label{eq:SF_MS_renormalized}
\end{equation}

For the quasi-SF, at leading power in the large-$v^z$ expansion, we obtain
\begin{equation}
   \begin{aligned}
      \widetilde S_{\overline{\mathrm{MS}}}(\xi,\mu) & =  \delta(\xi - \widetilde{k}) + \frac{\alpha_s C_F}{2\pi} \Bigg\{ \Bigg[ \frac{\theta(\xi - \widetilde{k})}{\xi - \widetilde{k}} \left( 2\ln \frac{2v^z(\xi - \widetilde{k})}{\mu} - 1 \right)  + \frac{\theta(\widetilde{k} - \xi)}{\widetilde{k} - \xi} \left( 2\ln \frac{2\mu v^z}{\widetilde{k} - \xi} - 3 \right) \Bigg]_{\oplus} \\
      &  + \delta(\xi - \widetilde{k}) \Bigg[ \left( \ln \frac{a - 1}{4(v^z)^2} + 2 \right) \ln \frac{\mu^2}{(a - 1)\xi^2} + \frac{1}{2}\ln^2[4(v^z)^2] - \ln[4(a - 1)(v^z)^2] + \frac{\pi^2}{6} + 2 \Bigg] \Bigg\}\,.
   \end{aligned} \label{eq:quasiSF_MS_renormalized}
\end{equation}

One finds that the ultraviolet divergences have been removed. Inserting these two renormalized SFs into Eq.~\eqref{eq:matching_one_loop}, we obtain
\begin{equation}
    \begin{aligned}
        & H_{\overline{\mathrm{MS}}/\overline{\mathrm{MS}}}(\xi,\omega,v^z,\mu) = \delta(\xi-\omega)  \\
        & \qquad + \frac{\alpha_s C_F}{2\pi} \Bigg\{ \left[  \frac{\theta(\xi-\omega)}{\xi-\omega} \left( 2\ln\frac{2v^z(\xi-\omega)}{\mu} - 1 \right) + \frac{\theta(\omega-\xi)}{\omega-\xi} \left( 2\ln \frac{2v^z(\omega-\xi)}{\mu} - 1 \right)  \right]_\oplus \\
        & \qquad +\delta(\xi-\omega) \left[ \frac{1}{2}\ln^2\frac{\mu^2}{4(a-1)\xi^2(v^z)^2} + \ln\frac{\mu^2}{4(a-1)\xi^2(v^z)^2} + \frac{1}{2}\ln^2(a-1) + 2+\frac{\pi^2}{4} \right]\Bigg\}\,.
    \end{aligned}\label{eq:matching_kernel_MS}
\end{equation}

The matching coefficient is ultraviolet finite and independent of the external-state infrared regulator $v\cdot k$. Its dependence on the boost $v^z$ and renormalization scale $\mu$ is logarithmic. The nonlocal correction has support on both sides of $\xi=\omega$, with the same functional dependence on $|\xi-\omega|$ away from the endpoint.

%%%%%%%%%%%%%%%%%%%%
\subsection{RI/MOM-to-$\overline{\rm MS}$ matching}
\label{sec:RI_renormalization}
%%%%%%%%%%%%%%%%%%%%

Keeping the light-cone SF in $\overline{\mathrm{MS}}$, we calculate the quasi-SF subtraction terms defined in Eq.~\eqref{eq:CTdef}. At the reference point, Eq.~\eqref{eq:reference} gives
\begin{equation}
   \begin{aligned}
       v\cdot k_R = v^0 k_R^0-v^z k_R^z=k_R^z\left(v^0\sqrt{1+r_R}-v^z\right)\,.
   \end{aligned} \label{eq:reference_energy}
\end{equation}
We expand these terms at large $v^z$, holding $k_R^z$ and $r_R$ fixed. The shifted arguments are specified by Eq.~\eqref{eq:RIshiftkernel}.

The heavy-quark sail diagram in Fig.~\ref{fig:feynman}(a) gives 
\begin{equation}
    \begin{aligned}
        & \widetilde{S}^{(1,a)}_{\mathrm{B,CT}}\left(\xi+\widetilde{k}_R-\widetilde{k},\{\widetilde\mu\}\right) \\
        & \qquad =\frac{\alpha_s C_F}{2\pi} \Bigg\{ \Bigg[ \frac{\theta(\xi - \widetilde{k})}{\xi - \widetilde{k}} \left( -\frac{1}{\epsilon} + 2\ln \frac{2v^z(\xi - \widetilde{k})}{\mu} + 1 \right) + \frac{\theta(\widetilde{k} - \xi)}{\widetilde{k} - \xi} \left( \frac{1}{\epsilon} + 2\ln \frac{\mu}{k_R^z(1 - \sqrt{1 + r_R})} \right) \Bigg]_{\oplus} \\
        & \qquad  + \delta(\xi - \widetilde{k}) \Bigg[ \ln \frac{a - 1}{4(v^z)^2} \left( \frac{1}{\epsilon} + 2\ln \frac{\mu}{k_R^z(1 - \sqrt{1 + r_R})} \right) + \frac{1}{2}\ln^2[4(v^z)^2] \\
        & \qquad + \ln^2 \frac{2v^z\xi}{k_R^z(1 - \sqrt{1 + r_R})}  - \frac{1}{\epsilon} + \ln \frac{8(v^z)^3 k_R^z(1 - \sqrt{1 + r_R})\xi}{\mu^2} + \frac{\pi^2}{2} \Bigg] \Bigg\}\,.
    \end{aligned}
\end{equation}

The Wilson-line self-energy diagram in Fig.~\ref{fig:feynman}(b) gives 
\begin{equation}
    \begin{aligned}
        \widetilde{S}^{(1,b)}_{\mathrm{B,CT}}\left(\xi+\widetilde{k}_R-\widetilde{k},\{\widetilde\mu\}\right) =\frac{\alpha_s C_F}{4\pi} \Bigg\{ -3 \left[ \frac{\theta(\xi - \widetilde{k})}{\xi - \widetilde{k}} + \frac{\theta(\widetilde{k} - \xi)}{\widetilde{k} - \xi} \right]_{\oplus} + \delta(\xi - \widetilde{k}) \left[ \frac{3}{\epsilon} + 4 + 3 \ln \frac{\mu^2}{4(a - 1)(v^z)^2\xi^2} \right] \Bigg\}\,.
    \end{aligned}\label{eq:RI_CT_Wilson}
\end{equation}
Since the external residual momentum enters this topology only through the Fourier phases, the shifted reference contribution equals the bare contribution and cancels it in the RI/MOM subtraction.

The nonlocal box diagram in Fig.~\ref{fig:feynman}(c) gives
\begin{equation}
    \begin{aligned}[b]
        \widetilde{S}^{(1,c)}_{\mathrm{B,CT}}\left(\xi+\widetilde{k}_R-\widetilde{k},\{\widetilde\mu\}\right) = -\frac{\alpha_s C_F}{4\pi} \Bigg\{ \Bigg[ \frac{\theta(\xi - \widetilde{k})}{\xi - \widetilde{k}} \Bigg]_{\oplus} + \delta(\xi - \widetilde{k}) \Bigg[ \ln \frac{2v^z\xi}{k_R^z(1 - \sqrt{1 + r_R})} - 1 \Bigg] \Bigg\}\,.
    \end{aligned} \label{eq:RI_CT_box}
\end{equation} 

The local vertex diagram in Fig.~\ref{fig:feynman}(d) gives
\begin{equation}
    \begin{aligned}[b]
        \widetilde{S}^{(1,d)}_{\mathrm{B,CT}}\left(\xi+\widetilde{k}_R-\widetilde{k},\{\widetilde\mu\}\right)=-\frac{\alpha_s C_F}{4\pi} \delta(\xi - \widetilde{k}) \Bigg[ \frac{3}{\epsilon} + 3\ln\frac{\mu^2}{4(v^z)^2(k_R^z)^2(1 - \sqrt{1 + r_R})^2} - 2 \Bigg]\,.
    \end{aligned}
\end{equation}

Combining the reference contributions with the local minus sign in Eq.~\eqref{eq:bareassembly}, and applying Eq.~\eqref{eq:RIshiftkernel}, we obtain
\begin{equation}
    \begin{aligned}
         & \widetilde{\mathcal Z}_{\mathrm{RI}}(\xi,\xi';v^z,\mu,\{\widetilde\mu\}) =\delta(\xi-\xi')\\
         & \qquad + \frac{\alpha_s C_F}{2\pi} \Bigg\{ \Bigg[ \frac{\theta(\xi - \xi')}{\xi - \xi'} \left( -\frac{1}{\epsilon} + 2\ln\frac{2v^z(\xi - \xi')}{\mu} - 1 \right)  + \frac{\theta(\xi' - \xi)}{\xi' - \xi} \left( \frac{1}{\epsilon} + 2\ln\frac{\mu}{k_R^z(1 - \sqrt{1 + r_R})} - \frac{3}{2} \right) \Bigg]_{\oplus} \\
         & \qquad + \delta(\xi - \xi') \Bigg[ \left( \ln\frac{a - 1}{4(v^z)^2} + 2 \right) \left( \frac{1}{\epsilon} + 2\ln\frac{\mu}{k_R^z(1 - \sqrt{1 + r_R})} \right) + \frac{1}{2}\ln^2[4(v^z)^2] + \ln^2\frac{2v^z\xi}{k_R^z(1 - \sqrt{1 + r_R})} \\
         & \qquad - \frac{1}{2}\ln[4(v^z)^2] - \frac{5}{2}\ln\frac{2v^z\xi}{k_R^z(1 - \sqrt{1 + r_R})} - \frac{3}{2}\ln(a - 1) + \frac{\pi^2}{2} + \frac{3}{2} \Bigg] \Bigg\}\,.
    \end{aligned}\label{eq:quasi_RI_Z}
\end{equation}
The pole terms agree with Eq.~\eqref{eq:quasi_MS_Z}; the additional finite terms are fixed by the RI/MOM reference condition. Substituting the one-loop part of this kernel into Eq.~\eqref{eq:commonsubtraction}, we obtain the renormalized quasi-SF through one loop at the stated accuracy:
\begin{equation}
   \begin{aligned}
      \widetilde S_{\mathrm{RI}}(\xi,\{\widetilde\mu\})&=\delta(\xi - \widetilde{k}) + \frac{\alpha_s C_F}{2\pi} \Bigg\{ \Bigg[ \frac{\theta(\widetilde{k} - \xi)}{\widetilde{k} - \xi} \left( 2\ln \frac{2v^z k_R^z(1 - \sqrt{1 + r_R})}{\widetilde{k} - \xi} - \frac{3}{2} \right) \Bigg]_{\oplus} \\
      & + \delta(\xi - \widetilde{k}) \Bigg[ -\ln^2 \frac{2v^z k_R^z(1 - \sqrt{1 + r_R})}{(a - 1)\xi} + \frac{3}{2}\ln \frac{2v^z k_R^z(1 - \sqrt{1 + r_R})}{(a - 1)\xi} + \frac{1}{2} - \frac{\pi^2}{3} \Bigg] \Bigg\}\,.
   \end{aligned}\label{eq:quasiSF_RI_renormalized}
\end{equation}
Substituting its one-loop term and that of the $\overline{\mathrm{MS}}$ light-cone SF Eq.~\eqref{eq:SF_MS_renormalized} into Eq.~\eqref{eq:matching_one_loop}, we obtain the matching coefficient through one loop,
\begin{equation}
    \begin{aligned}
       & H_{\mathrm{RI}/\overline{\mathrm{MS}}}(\xi,\omega;v^z,\mu,k_R^z,r_R)=\delta(\xi-\omega)+ \frac{\alpha_s C_F}{2\pi} \Bigg\{ \Bigg[ \frac{\theta(\omega - \xi)}{\omega - \xi} \left( 2\ln \frac{2v^z k_R^z(1 - \sqrt{1 + r_R})(\omega - \xi)}{\mu^2} + \frac{1}{2} \right) \Bigg]_{\oplus} \\
       & \qquad + \delta(\xi - \omega) \Bigg[ -\ln^2 \frac{2v^z k_R^z(1 - \sqrt{1 + r_R})}{(a - 1)\xi} + \frac{1}{2}\ln^2 \frac{(a - 1)^2\xi^2}{\mu^2} \\
       & \qquad+ 2\ln \frac{2v^z k_R^z(1 - \sqrt{1 + r_R})}{\mu} - \frac{1}{2}\ln \frac{2v^z k_R^z(1 - \sqrt{1 + r_R})}{(a - 1)\xi} + \frac{1}{2} - \frac{\pi^2}{4} \Bigg] \Bigg\}\,.
    \end{aligned}\label{eq:matching_kernel_RI}
\end{equation}

Comparing Eqs.~(\ref{eq:matching_kernel_MS}) and (\ref{eq:matching_kernel_RI}), we find that the difference between the two matching coefficients arises entirely from the different renormalization schemes used for the quasi-SF. Since both calculations involve the same bare quasi-SF and the same $\overline{\mathrm{MS}}$ light-cone SF, their common contributions cancel, leaving the difference between the quasi-SF renormalization kernels:
\begin{equation}
    \begin{aligned}
        H_{\overline{\mathrm{MS}}/\overline{\mathrm{MS}}}^{(1)}(\xi,\omega)-H_{\mathrm{RI}/\overline{\mathrm{MS}}}^{(1)}(\xi,\omega) = \left.\left[\widetilde{\mathcal Z}_{\mathrm{RI}}^{(1)}(\xi,\xi')-\widetilde{\mathcal Z}_{\overline{\mathrm{MS}}}^{(1)}(\xi,\xi')\right]\right|_{\xi'\to\omega}\,.
    \end{aligned} \label{eq:matching_scheme_difference}
\end{equation}
The ultraviolet poles cancel in the difference of the renormalization kernels, leaving a finite contribution determined by the RI/MOM subtraction condition.

The explicit matching coefficient in Eq.~\eqref{eq:matching_kernel_RI} has a nonlocal correction only for $\xi<\omega$; the nonlocal correction vanishes for $\xi>\omega$. This behavior results from the finite RI/MOM subtraction, which cancels the corresponding nonlocal contribution to the $\overline{\mathrm{MS}}$ matching coefficient in the region $\xi>\omega$.

Notably, this momentum-region structure differs from that found in the $B$-meson LCDA calculation of Ref.~\cite{Xu:2022guw}. In that calculation, the light-quark propagator $1/(q+k_R)^2$ introduces the absolute-value structures $|q^z|$ and $|q^z+k_R^z|$ into the loop integrals. The latter changes form at $q^z=-k_R^z$, producing an additional boundary at $\xi=\omega-k_R^z/v^z$. In the present SF calculation, the loop integrals instead involve HQET heavy-quark propagators, and no corresponding additional boundary appears at one loop and leading power.

Although the matching coefficients differ because of the scheme-dependent finite subtraction on the quasi-SF side, both consistently relate their respective renormalized quasi-SFs to the same $\overline{\mathrm{MS}}$ light-cone SF. The extracted light-cone SF is therefore independent of the quasi-SF renormalization scheme, up to higher-order perturbative and power corrections.

\end{widetext}
%%%%%%%%%%%%%%%%%%%%%	
\section{Summary and outlook}
\label{sec:summary}
%%%%%%%%%%%%%%%%%%%%%

We have established the direct perturbative matching between the quasi-SF and the light-cone SF, both defined within HQET. We have calculated the one-loop renormalization kernels and matching coefficients for the quasi-SF in the $\overline{\mathrm{MS}}$ and RI/MOM schemes, while keeping the light-cone SF in the $\overline{\mathrm{MS}}$ scheme. The resulting matching coefficients are ultraviolet finite and independent of the external-state infrared regulator $v\cdot k$. In particular, the RI/MOM matching coefficient exhibits a scheme dependent momentum region structure, with its nonlocal contribution restricted to $\xi<\omega$ at one loop and leading power in the large-$v^z$ expansion.
	
From a Wilsonian effective field theory perspective, the direct HQET approach provides a natural framework for lattice calculations in which the heavy quark mass scale is integrated out before evaluating the nonperturbative matrix elements. Compared with the two-step construction, it avoids introducing an intermediate QCD light-cone SF and the associated QCD-to-HQET matching step. Our results provide the necessary perturbative input for a future lattice HQET determination of the $B$-meson shape function.

%%%%%%%%%%%%%%%%%%%%%
\section*{Acknowledgements}
%%%%%%%%%%%%%%%%%%%%%
The authors would like to thank Prof. Wei Wang for inspiring discussions. This work is supported in part by the National Natural Science Foundation of China under Grant No. 12675117, 12475098, 12105247, 12335003, 12247101, and by the Fundamental Research Funds for the Central Universities under No. lzujbky-2023-stlt01, lzujbky-2024-oy02, lzujbky-2025-jdzx07 and lzujbky-2025-eyt01, the Scientific Research Innovation Capability Support Project for Young Faculty under Grant No. ZYGXQNJSKYCXNLZCXMP2, the Natural Science Foundation of Gansu Province (No.25JRRA799), and the ‘111 Center’ under Grant No. B20063. J.X is supported in part by the Key Laboratory for Particle Astrophysics and Cosmology, Ministry of Education (MoE). S.M.H is supported by the Fundamental Research Funds for the Central Universities under No. lzujbky-2023-stlt01.

\end{document}